\documentstyle[amsfonts,11pt]{article}

\def\ber#1#2{\begin{equation}\begin{array}{#1}\displaystyle{#2}}
\def\ber#1{\begin{equation}\begin{array}{#1}\displaystyle}
\def\bernn#1#2{$$\begin{array}{#1}\displaystyle{#2}}

\def\eer#1{\end{array}\label{#1}\end{equation}}
\def\eernn{\end{array}$$}
\def\r#1#2{\noindent\hbox{\hbox to 24 pt{\hfil[#1]~}%
\vtop{\hsize = 12.5 truecm\noindent#2}}\vskip 5 pt\vfil}
\def\chap#1#2#3{\noindent\hbox{\hbox to 1.5 truecm{\hfil#1}%
\hbox to 14 truecm{~#2\leaders\hbox to 0.5 em{\hfil.\hfil}\hfill#3}}\par}
\def\cchap#1#2#3#4{\noindent\hbox{\hbox to 1.5 truecm{\hfil#1}%
\hbox to 14 truecm{~#2\hfil}}\par
\noindent\hbox{\hskip 1.5 truecm%
\hbox to 14 truecm{~#3\leaders\hbox to 0.5 em{\hfil.\hfil}\hfill#4}}\par}
\def\bbt{\bibitem}
\def\be{\begin{equation}}
\def\en{\end{equation}}
\def\ber{\begin{eqnarray}}
\def\enr{\end{eqnarray}}
\def\nmb{ \nonumber\\}
\def\d{\partial}
\def\rbr{\rbrack}
\def\lbr{\lbrack}
\def\rbrc{\rbrace}
\def\lbrc{\lbrace}
\def\ov{\over }
\def\tld{\tilde}
\def\brv{\breve}

\def\MTR{Manin triple }

\def\DLG{double Lie group }
\def\Tta{\Theta}
\def\sgm{\sigma}
\def\Sgm{\Sigma}
\def\al{\alpha}
\def\bt{\beta}
\def\gm{\gamma}
\def\im{\imath}

\def\Lm{\Lambda}

\def\et{\eta}
\def\eps{\epsilon}
\def\dlt{\delta}
\def\Dl{\Delta}
\def\Gm{\Gamma}
\def\Sgm{\Sigma}

\begin{document}
\rightline{Landau Tmp/11/98.}
\rightline{November 1998}
\vskip 2 true cm
\centerline{\bf ON THE QUANTUM POISSON-LIE T-DUALITY AND MIRROR SYMMETRY.}
\vskip 2.5 true cm
\centerline{\bf S. E. Parkhomenko}
\centerline{Landau Institute for Theoretical Physics}
\centerline{142432 Chernogolovka,Russia}
\vskip 0.5 true cm
\centerline{spark@itp.ac.ru}
\vskip 1 true cm
\centerline{\bf Abstract}
\vskip 0.5 true cm

 Poisson-Lie $T$-duality in quantum $N=2$ superconformal
WZNW models is considered. The Poisson-Lie $T$-duality
transformation rules of the super-Kac-Moody algebra currents
are found from the conjecture that, as in the classical case,
the quantum Poisson-Lie $T$-duality transformation is given by an
automorphism which interchanges the isotropic subalgebras of the 
underlying Manin triple in one of the chirality sectors of the model. 
It is shown that quantum Poisson-Lie $T$-duality
acts on the $N=2$ super-Virasoro algebra generators of the
quantum models as a mirror symmetry acts:
in one of the chirality sectors it is a trivial transformation 
while in another
chirality sector it changes the sign of the $U(1)$ current and
interchanges the spin-3/2 currents. A generalization of Poisson-Lie
$T$-duality for the quantum Kazama-Suzuki models is proposed.
It is shown that quantum Poisson-Lie $T$-duality acts in these
models as a mirror symmetry also.



\smallskip
\vskip 10pt
\centerline{\bf Introduction.}

 Target-space ($T$) dualities in superstring theory relate backgrounds
with different geometries and are symmetries of the underlying conformal 
field theory ~\cite{GiPR}, ~\cite{AlvG}.

 The mirror symmetry ~\cite{MS} discovered in superstring theory
is a special type of $T$-duality. At the level of conformal
field theory it can be formulated as an isomorphism between
two theories, amounting to a change of sign of the $U(1)$
generator and an interchange of the spin-3/2 generators of the leftmoving 
(or rightmoving) $N=2$ superconformal algebra.

 Mirror symmetry has mostly been studied in the context of
Calabi-Yau superstring compactification. 
Important progress has been achieved in this direction in the last few years, 
based on the ideas of toric geometry ~\cite{MoPl1}. In particular, in Ref.
~\cite{BatBor} toric geometry mirror pair construction was proposed.
Though it seems quite certain that pairs of Calabi-Yau
manifolds constructed by these methods are mirror, one needs to show that the
proposed pairs correspond to isomorphic conformal field theories, to prove that 
they are indeed mirror. Progress in this direction was made in ~\cite{MoPl2},
but a complete arguments has yet to be carried out.
In fact, the only rigorously established example of mirror symmetry,
the Green-Plesser construction ~\cite{GrPles}, is based on the tensor
products of the $N=2$ minimal models ~\cite{MM}. For a review of 
mirror symmetry and toric geometry methods in Calabi-Yau superstring 
compactifications see the lectures of B. Greene ~\cite{Grl}.

 Recently, A. Strominger, S-T. Yau, E. Zaslow 
~\cite{StYauZ} related mirror symmetry in superstring theory 
to the quantum Abelian $T$-duality in fibers of 
toricaly fibrated Calabi-Yau manifolds. 

 The Poisson-Lie (PL) $T$-duality,
recently discovered by C. Klimcik and P. Severa in their excellent
work ~\cite{KlimS1}, is a
generalization of Abelian and non-Abelian $T$-dualities
~\cite{OsQ}-~\cite{FrJ}.
This generalized duality is associated with two groups
forming a Drinfeld double ~\cite{Drinf1}, and the duality transformation
exchanges their roles. Many aspects of their ideas have been developed 
in Refs. ~\cite{KlimS2}-~\cite{PLM}.
In particular, in ~\cite{PLM} it was shown that PL $T$-duality in the 
classical $N=2$ superconformal WZNW (SWZNW) and Kazama-Suzuki models 
is a mirror duality. It is reasonable to expect that
PL $T$-duality in the quantum versions of these models
will be a mirror duality also. Moreover, it is tempting to conjecture 
that PL $T$-duality
is an adequate geometric structure underlying mirror symmetry
in superstring theory. Motivated by this we propose a quantization of 
PL $T$-duality transformations in the $N=2$ SWZNW and Kazama-Suzuki models.

 Quantum equivalence among PL $T$-duality related
$\sgm$-models was studied perturbatively in ~\cite{BFMPS} and
~\cite{Sfet2}, and it was shown that PL dualizability is
compatible with renormalization at 1 loop. 
In particular it was shown in ~\cite{Sfet2} that 1-loop beta functions for 
the coupling and the parameters in the two simplest examples of 
PL $T$-duality related models are equivalent. This allows us to suggest that
their equivalence extends beyond the classical level with appropriate 
quantum modification of PL $T$-duality transformations rules. 

 In the present note the PL $T$-duality transformation rules of the fields in 
quantum $N=2$ SWZNW models will be found
starting from the conjecture that as in the classical case, 
quantum $N=2$ SWZNW models are PL self-dual and the PL $T$-duality 
transformation 
is given by an automorphism of the super-Kac-Moody algebra
in the rightmoving sector. 
Then we obtain PL $T$-duality transformation rules using 
the Knizhnik-Zamolodchikov equation,
Ward identities and a quantum version of the classical formula which 
relates the generators of rightmoving super-Kac-Moody algebra to its
PL $T$-duality transformed. We show that the generators
of the $N=2$ super-Virasoro algebras transform under
PL $T$-duality like a mirror duality:
the $U(1)$ current changes sign and the spin-3/2 currents permute.
Thus, the results are in
agreement with the conjecture proposed in ~\cite{GiWi}
that mirror symmetry can be related to a gauge symmetry (automorphism)
of the self-dual points of the moduli space of the $N=2$ 
superconformal field theories (SCFTs) 
(for the N=0 version of this conjecture see ~\cite{GiKir}). Then 
we consider quantum PL $T$-duality in the Kazama-Suzuki models 
and propose a natural 
generalization of the quantum PL $T$-duality transformation. 
We show that as in the SWZNW models quantum PL $T$-duality in the 
Kazama-Suzuki models is a mirror duality also.

 The structure of the paper is as follows:
In section 1 we briefly review PL $T$-duality in the classical
$N=2$ SWZNW model following ~\cite{PLM}.
In section 2 we describe Manin triple construction of the quantum
$N=2$ SWZNW models on the compact groups and obtain
the PL $T$-duality transformation rules of the quantum fields.  
We show that PL $T$-duality transformation is given by 
an automorphism of the underlying Manin triple which permutes
isotropic subalgebras of the triple. Then we obtain
transformation rules of the rightmoving $N=2$ super-Virasoro algebra 
generators.
In section 3 we present the Manin triple construction of
the Kazama-Suzuki models. We show that they can be described
as $(Manin \ triple)/(Manin \ subtriple)$-cosets. We define quantum 
PL $T$-duality transformation in the Kazama-Suzuki models as the subset of 
the transformations of the numerator triple which
stabilizes the denominator subtriple. Then we easily find transformation rules
of the rightmoving $N=2$ super-Virasoro algebra generators of the coset. 
At the end of the section PL $T$-duality in the $N=2$ minimal models 
considered briefly as an example.

\vskip 10pt
\centerline{\bf1. Poisson-Lie $T$-duality and Mirror Symmetry in}
\centerline{\bf the classical $N=2$ superconformal WZNW models.}

 In this section we briefly review PL $T$-duality in the 
classical $N=2$ SWZNW models, following  ~\cite{PLC, PLM}.

 We parameterize the super world-sheet by introducing the light cone
coordinates
$z_{\pm}$ and Grassman coordinates $\Tta_{\pm}$ (we use the N=1
superfield formalism).
The generators of the supersymmetry and covariant
derivatives satisfying the standard relations are given by

\be
Q_{\mp}= {\d \ov \d\Tta_{\pm}}+\im \Tta_{\pm}\d_{\mp},\
D_{\mp}= {\d \ov \d\Tta_{\pm}}-\im \Tta_{\pm}\d_{\mp}.
\label{1}
\en
The superfield of the N=2 SWZNW model
\be
G= g+ \im \Tta_{-}\psi_{+}+ \im \Tta_{+}\psi_{-}+
   \im \Tta_{-}\Tta_{+}F  \label{2}
\en
takes values in a compact Lie group ${\bf G}$ so that
its Lie algebra ${\bf g}$
is endowed with an ad-invariant nondegenerate inner
product $<,>$.
The action of the model is given by
\ber
S_{SWZ}= \int d^{2}x d^{2} \Tta(<G^{-1}D_{+}G,G^{-1}D_{-}G>)   \nmb
         -\int d^{2}x d^{2}\Tta dt
          <G^{-1}\frac{\d G}{\d t},\lbrc G^{-1}D_{-}G,G^{-1}D_{+}G\rbrc>
\label{3}
\enr
and possesses manifest $N=1$ superconformal and super-Kac-Moody
symmetries ~\cite{swzw}:
\ber
\dlt_{a_{+}}G(z_{+},x_{-},\Tta_{+},\Tta_{-})=
a_{+}(z_{-},\Tta_{+})G(z_{+},z_{-},\Tta_{+},\Tta_{-}), \nmb
\dlt_{a_{-}}G(z_{+},z_{-},\Tta_{+},\Tta_{-})=
-G(z_{+},z_{-},\Tta_{+},\Tta_{-})a_{-}(z_{+},\Tta_{-}),
\label{km}
\enr
\ber
G^{-1}\dlt_{\eps_{+}}G=(G^{-1}\eps_{+}(z_{-})Q_{+}G), \nmb
\dlt_{\eps_{-}}GG^{-1}=\eps_{-}(z_{+})Q_{-}GG^{-1},
\label{su}
\enr
where $a_{\pm}$ are ${\bf g}$-valued superfields.

 An additional ingredient demanded by the $N=2$ superconformal
symmetry is a complex structure $J$
on the finite-dimensional Lie algebra of the model which is
skew-symmetric with respect to the inner product $<,>$ ~\cite{QFR3,QFR,QFR2}.
That is, we should demand that the following equations be satisfied
on ${\bf g}$:
\ber
J^{2}=-1, \nmb
<Jx,y>+<x,Jy>=0,  \nmb
\lbr Jx,Jy \rbr-J\lbr Jx,y \rbr-J\lbr x,Jy \rbr=\lbr x,y \rbr \label{4}
\enr
for any elements $x, y$ in ${\bf g}$.
It is clear that the corresponding Lie group is a complex
manifold with left (or right) invariant complex structure.
In the following we shall denote  the real Lie group and the real
Lie algebra with the complex structure satisfying (\ref{4}) by the
pairs $({\bf G},J)$ and $({\bf g},J)$ respectively.

 The complex structure $J$ on the Lie algebra defines the
second supersymmetry transformation ~\cite{QFR2}
\ber
(G^{-1}\dlt_{\et_{+}}G)^{a}=\et_{+}(z_{-})(J_{l})^{a}_{b}(G^{-1}D_{+}G)^{b}, \nmb
(\dlt_{\et_{-}}GG^{-1})^{a}=\et_{-}(z_{+})(J_{r})^{a}_{b}(D_{-}GG^{-1})^{b},
\label{Rsu}
\enr
where $J_{l}, J_{r}$ are the left invariant and right invariant
complex structures on ${\bf G}$ which correspond to the
complex structure $J$.

 The notion of Manin triple is closely related to a complex structure
on a Lie algebra. By definition ~\cite{Drinf1},
a \MTR $({\bf g},{\bf g_{+}},{\bf g_{-}})$
consists of a Lie algebra ${\bf g}$ with nondegenerate invariant inner
product $<,>$ and isotropic Lie subalgebras ${\bf g_{\pm}}$ such that
${\bf g}={\bf g_{+}}\oplus {\bf g_{-}}$ as a vector space.

 With each pair $({\bf g},J)$ one can associate the complex \MTR
$({\bf g^{\Bbb C}},{\bf g_{+}},{\bf g_{-}})$, where
${\bf g^{\Bbb C}}$ is the complexification of ${\bf g}$ and
${\bf g_{\pm}}$ are ${\pm}\im$ eigenspaces of $J$.
Moreover, it can be proved  that there exists a one-to-one
correspondence between a complex Manin triple endowed with an anti-linear
involution which conjugates isotropic subalgebras
$\tau: {\bf g_{\pm}}\to
{\bf g_{\mp}}$ and a real Lie
algebra endowed with an $ad$-invariant nondegenerate inner product $<,>$
and complex structure $J$ which is skew-symmetric with respect
to $<,>$ ~\cite{QFR3}.
The conjugation can be used to extract a real form from a complex
Manin triple.

 Now we have to consider some geometric properties of the $N=2$ SWZNW
models closely related to the existence of complex structures
on the groups. We shall follow ~\cite{PLC}.

 Let us fix some compact Lie group with the left invariant complex
structure $({\bf G}, J)$ and consider its Lie algebra with
the complex structure $({\bf g}, J)$.
The complexification ${\bf g^{\Bbb C}}$ of ${\bf g}$ has the Manin triple
structure $({\bf g^{\Bbb C}},{\bf g_{+}},{\bf g_{-}})$. The Lie group version
of this triple  is the \DLG $({\bf G^{\Bbb C}},{\bf G_{+}},{\bf G_{-}})$
~\cite{SemTian,AlMal,LuW}, where the exponential subgroups
${\bf G_{\pm}}$ correspond to the Lie algebras
${\bf g_{\pm}}$. The real Lie group ${\bf G}$ is extracted
from its complexification with the help of conjugation $\tau$
(it will be assumed in the following that $\tau$ is the hermitian
conjugation)
\be
{\bf G}= \lbrc g\in {\bf G^{\Bbb C}}|\tau (g)=g^{-1}\rbrc       \label{rf}
\en
 Each element $g\in {\bf G^{\Bbb C}}$ from the
vicinity ${\bf G_{1}}$ of the unit element from ${\bf G^{\Bbb C}}$
admits two decompositions:
\be
g= g_{+}g^{-1}_{-}= {\tld g}_{-}{\tld g}^{-1}_{+}.  \label{13}
\en
Taking into account (\ref{rf}) and (\ref{13}) we conclude that the
element $g$ ($g\in {\bf G_{1}}$) belongs to ${\bf G}$ iff
\be
\tau (g_{\pm})= {\tld g}^{-1}_{\mp}      \label{13u}
\en
These equations mean that we can parameterize the elements of
\be
{\bf C_{1}}\equiv {\bf G_{1}}\cap {\bf G} \label{13cl}
\en
by the elements of the complex group ${\bf G}_{+}$ (or ${\bf G}_{-}$),
i.e., we can introduce complex coordinates (they are just matrix elements
of $g_{+}$ (or $g_{-}$)) in the strat ${\bf C_{1}}$.

 To generalize (\ref{13}), (\ref{13u}) one has to consider the
set $W$ (which we shall assume in the following to be discrete and
finite) of classes ${\bf G_{+}}\backslash {\bf G^{\Bbb C}}/ {\bf G_{-}}$
and chose a representative $w$ for each class $[w]\in W$.
It gives us the stratification
of ${\bf G^{\Bbb C}}$ ~\cite{AlMal}:
\be
{\bf G^{\Bbb C}}= \bigcup_{[w]\in W} {\bf G_{+}}w{\bf G_{-}}=
         \bigcup_{[w]\in W} {\bf G_{w}}  \label{17+}
\en
There is a second stratification:
\be
{\bf G^{\Bbb C}}= \bigcup_{[w]\in W} {\bf G_{-}}w{\bf G_{+}}=
         \bigcup_{[w]\in W} {\bf G^{w}}  \label{17-}
\en
We shall assume, in the following, that the representatives $w$
have been chosen to be unitary:
\be
\tau (w)=w^{-1}                        \label{17w}
\en
It allows us to generalize (\ref{13}) as follows:
\be
g= wg_{+}g^{-1}_{-}= w{\tld g}_{-}{\tld g}^{-1}_{+},    \label{18t}
\en
where
\be
g_{+}\in {\bf G^{w}_{+}},\
{\tld g}_{-}\in {\bf G^{w}_{-}} \label{037}
\en
and
\be
{\bf G^{w}_{+}}= {\bf G_{+}}\cap w^{-1}{\bf G_{+}}w, \
{\bf G^{w}_{-}}= {\bf G_{-}}\cap w^{-1}{\bf G_{-}}w.
\label{038}
\en

 In order for the element $g$ to belong to the real group ${\bf G}$
the elements $g_{\pm}, {\tld g}_{\pm}$ from (\ref{18t})
must satisfy (\ref{13u}).
Thus, the formulas (\ref{13u}, \ref{18t})
define the mapping
\be
\phi^{+}_{w}: {\bf G^{w}_{+}}\to {\bf C_{w}}\equiv {\bf G_{w}}\cap {\bf G}
\label{037m}
\en
In a similar way one can define the mapping
\be
\phi^{-}_{w}: {\bf G^{w}_{-}}\to {\bf C_{w}}\equiv {\bf G_{w}}\cap {\bf G}.
\label{037m-}
\en

 In ~\cite{PLC, PLM} the following statements were proved:
\begin{itemize}
\item
the mappings (\ref{037m})
are holomorphic and define the natural (holomorphic) action of the complex 
group ${\bf G_{+}}$  on ${\bf G}$;
the set $W$ parameterizes the ${\bf G_{+}}$-orbits ${\bf C_{w}}$.

\item
the $({\bf G},J)$-SWZNW model admits PL symmetry ~\cite{KlimS1},
~\cite{PLSY}
with respect to ${\bf G_{+}}$-action, so that we may associate with
each extremal surface
$G_{+}(z_{+}, z_{-}, \Tta_{+}, \Tta_{-})\subset {\bf G_{+}}$,
of the model a mapping ("Noether charge")
$V_{-}(z_{+}, z_{-}, \Tta_{+}, \Tta_{-})$
from the super world-sheet into the group ${\bf G_{-}}$.
The pair $(G_{+}(z_{+}, z_{-}, \Tta_{+}, \Tta_{-}),
V_{-}(z_{+}, z_{-}, \Tta_{+}, \Tta_{-}))$
can be lifted into the the double ${\bf G^{\Bbb C}}$:
\ber
\Phi(z_{+}, z_{-}, \Tta_{+}, \Tta_{-})=
G_{+}(z_{+}, z_{-}, \Tta_{+}, \Tta_{-})
V_{-}(z_{+}, z_{-}, \Tta_{+}, \Tta_{-}).
\label{42}
\enr
Moreover, the surface (\ref{42}) can be rewritten in the form
\be
\Phi(z_{\pm}, \Tta_{\pm})=G(z_{\pm}, \Tta_{\pm})H^{-1}_{-}(z_{+}, \Tta_{-}),
\label{42.1}
\en
where $G(z_{\pm}, \Tta_{\pm})\subset {\bf G}$ is a solution of
the ${\bf G}$-SWZNW model and the superfield $H_{-}$ is given by the
solution of the equation
\be
H^{-1}_{-}D_{+}H_{-}=2(I_{+})^{-},
\label{Jr4}
\en
where $(I_{+})^{-}$ is ${\bf g_{-}}$-projection of the conservation
current $I_{+}=G^{-1}D_{+}G$ of the model.

\item
  With the appropriate modifications the above statements
are true also for the mappings (\ref{037m-}) and ${\bf G_{-}}$-action
on ${\bf G}$. Thus, one can represent the surface (\ref{42})
in the "dual" parameterization ~\cite{KlimS1}
\be
\Phi(z_{\pm}, \Tta_{\pm})=
\brv{G}(z_{\pm}, \Tta_{\pm})H^{-1}_{+}(z_{+}, \Tta_{-})
,\label{48}
\en
where $\brv{G}(z_{\pm}, \Tta_{\pm})$ is the dual
solution of the ${\bf G}$-SWZNW model and the superfield $H_{+}$
is given by the similar equation
\be
H^{-1}_{+}D_{+}H_{+}=2(\brv{I}_{+})^{+},
\label{Jr5}
\en
where $(\brv{I}_{+})^{+}$ is the ${\bf g_{+}}$-projection of
the dual conserved current
$\brv{I}_{+}\equiv \brv{G}^{-1}D_{+}\brv{G}$.

\item
 Under PL $T$-duality
\be
t: G(z_{\pm}, \Tta_{\pm}) \to \brv{G}(z_{\pm}, \Tta_{\pm})=
G(z_{\pm}, \Tta_{\pm})H(z_{+}, \Tta_{-}),
\label{tmap}
\en
where
\be
H\equiv H^{-1}_{-}H_{+},
\label{H}
\en
the conserved rightmoving current $I_{+}$ transforms as
\be
t: (I_{+})^{-}\to (\brv{I}_{+})^{+}, \
   (I_{+})^{+}\to (\brv{I}_{+})^{-},
\label{Imap}
\en
while the conserved leftmoving current $I_{-}\equiv D_{-}GG^{-1}$ 
transforms identically:
\be
t: (I_{-})^{\pm}\to (I_{-})^{\pm}.
\label{Imap-}
\en
Moreover, the classical rightmoving
$N=2$ super-Virasoro algebra maps under PL $T$-duality
as follows~\cite{PLM}:
\be
t: \Sgm^{\pm}\to \brv{\Sgm}^{\mp}, \
T\pm \im \d K \to \brv{T}\mp \im \d \brv{K},
\label{62}
\en
where $\Sgm^{\pm}$ are the spin 3/2- currents,
$T$ is the stress-energy tensor, and $K$ is the $U(1)$ current,
while the leftmoving $N=2$ super-Virasoro algebra maps identically.
Thus, PL $T$-duality in the classical $N=2$ SWZNW models is
a mirror duality.

\end{itemize}

\vskip 10pt
\centerline{\bf2. Poisson-Lie $T$-duality and Mirror Symmetry in}
\centerline{\bf the quantum $N=2$ superconformal WZNW models.}
 We start with the \MTR construction of the $N=2$ Virasoro algebra
generators of the quantum SWZNW model on the group $({\bf G}, J)$
~\cite{QFR3}-~\cite{GETZ}.

 Let us specify an orthonormal basis
\be 
\{E^{a}, E_{a}, a= 1,...,d\}
\label{Bas}
\en
in the \MTR $({\bf g^{\Bbb C}},{\bf g_{+}},{\bf g_{-}})$
so that $\{E^{a}\}$ is a basis in $g_{+}$, and $\{E_{a}\}$
is a basis in $g_{-}$. The commutation relations and Jacoby identity in 
this basis take the form
\ber
\lbr E^{a},E^{b}\rbr=f^{ab}_{c}E^{c},     \nmb
\lbr E_{a},E_{b}\rbr=f_{ab}^{c}E_{c},     \nmb
\lbr E^{a},E_{b}\rbr=f_{bc}^{a}E^{c}-f^{ac}_{b}E_{c},
\label{2.1}
\enr
\ber
f^{ab}_{d}f^{dc}_{e}+f^{bc}_{d}f^{da}_{e}+ f^{ca}_{d}f^{db}_{e}=0, \nmb
f_{ab}^{d}f_{dc}^{e}+f_{bc}^{d}f_{da}^{e}+ f_{ca}^{d}f_{db}^{e}=0, \nmb
f_{mc}^{a}f^{bm}_{d}-f_{md}^{a}f^{bm}_{c}-f_{mc}^{b}f^{am}_{d}+
f_{md}^{b}f^{am}_{c}= f_{cd}^{m}f^{ab}_{m}.
\label{2.2}
\enr
Let us introduce the matrices
\ber
B^{b}_{a}= f_{c}f^{cb}_{a}+f^{c}f_{ca}^{b}, \nmb
A^{b}_{a}= f_{ac}^{d}f^{bc}_{d}.
\label{2.3}
\enr
Let $j^{a}(z), j_{a}(z)$ be the generators of the affine Kac-Moody
algebra $\hat{g}^{\Bbb C}$, corresponding to the fixed
basis $\{E^{a}, E_{a}\}$, so that the currents $j^{a}$ generate
the subalgebra $\hat{g}_{+}$ and the currents $j_{a}$ generate
the subalgebra $\hat{g}_{-}$ (we shall omit in the following
the super-world-sheet indices ${\pm}$, keeping in mind
that we are in the rightmoving sector). The singular OPEs between these
currents are the following:
\ber
j^{a}(z)j^{b}(w)=-(z-w)^{-2}{1\ov 2}k(E^{a},E^{b})
          +(z-w)^{-1}f^{ab}_{c}j^{c}(w)+reg,   \nmb
j_{a}(z)j_{b}(w)=-(z-w)^{-2}{1\ov 2}k(E_{a},E_{b})
          +(z-w)^{-1}f_{ab}^{c}j_{c}(w)+reg,   \nmb
j^{a}(z)j_{b}(w)=-(z-w)^{-2}{1\over 2}
          (q\delta^{a}_{b}+k(E^{a},E_{b}))      \nmb
          +(z-w)^{-1}(f_{bc}^{a}j^{c}-f^{ac}_{b}j_{c})(w)+reg,
\label{2.4}
\enr
where $k(x,y)$ denotes the Killing form for the vectors $x,y$
of ${\bf g^{\Bbb C}}$.
Let $\psi^{a}(z), \psi_{a}(z)$ be free fermion currents which
have the following singular OPEs:
\be
 \psi^{a}(z)\psi_{b}(w)= -(z-w)^{-1}\delta^{a}_{b}+reg.
\label{2.5}
\en
Then the $N=2$ Virasoro superalgebra currents and the central charge
are given by
~\cite{QFR2}-~\cite{GETZ}
\ber
\Sgm^{+}={2\over \sqrt{q}}(\psi^{a}j_{a}+
          {1\over 2}f_{ab}^{c}:\psi^{a}\psi^{b}\psi_{c}:),   \nmb
\Sgm^{-}={2\over \sqrt{q}}(\psi_{a}j^{a}+
          {1\over 2}f^{ab}_{c}:\psi_{a}\psi_{b}\psi^{c}:),   \nmb
K= ({2B^{b}_{a}\over q} -\delta^{b}_{a}):\psi^{a}\psi_{b}:-
          {2\over q}(f_{c}j^{c}-f^{c}j_{c}),             \nmb
T=-{1\over q}:(j^{a}j_{a}+j_{a}j^{a}):-
         {1\ov 2} :(\d \psi^{a}\psi_{a}-\psi^{a}\d \psi_{a}):,
\label{2.6}
\enr
\be
 c= 3(d-{2A^{a}_{a}\over q}).
\label{2.7}
\en

The set of currents (\ref{2.6}) can be combined into the
superfields
\be
\Gm^{\pm}={1\over \sqrt{2}}\Sgm^{+}+\Tta (T\mp {1\over 2}\d K),
\label{2.8}
\en
so that the energy-momentum super-tensor is given by the sum
\be
\Gm={1\ov 2}(\Gm^{+}+\Gm^{-})=
-{1\over q}:<DI,I>:+{2\over 3q^{2}}:<I,:\lbrc I,I\rbrc:>:.
\label{2.9}
\en
Here $I$ denotes Lie algebra valued super-Kac-Moody currents
of the affine superalgebra $\hat{\bf g}$:
\ber
I\equiv I^{a}E_{a}+I_{a}E^{a}, \nmb
I^{a}=-\sqrt{q\ov 2}\psi^{a}+\Tta(j^{a}+
({1\over 2}f_{bc}^{a}:\psi^{b}\psi^{c}:+ f^{ab}_{c}:\psi_{b}\psi^{c}:)), \nmb
I_{a}=-\sqrt{q\ov 2}\psi_{a}+\Tta(j_{a}+
({1\over 2}f^{bc}_{a}:\psi_{b}\psi_{c}:+ f_{ab}^{c}:\psi^{b}\psi_{c}:)).
\label{2.10}
\enr

 We now propose a quantum version of the PL $T$-duality
transformation.
Perhaps the most comprehensive way to find PL $T$-duality transformation
rules for the quantum fields of the model is to quantize 
canonically the Sfetsos canonical transformations
for PL $T$-duality related $\sgm$-models ~\cite{Sfet1} and then  
define and solve 
the quantum version of the equations (\ref{Jr4}, \ref{Jr5}, \ref{H}). 
Though developing
of this approach for the $N=2$ superconformal field theory
is an important problem and worth solving, it is beyond our reach
at the present moment.

Instead we determine the quantum counterpart of the mapping (\ref{tmap}) as 
an automorphism of the operator algebra of the quantum fields, 
defined by right multiplication by the rightmoving
matrix-valued function $H(Z)$, which implies that $N=2$ SWZNW
model is PL self-dual.
We propose a very simple way to find the matrix elements of $H$ 
using super-Kac-Moody Ward identities and the Knizhnik-Zamolodchikov
equation.

 In the $N=1$ superfield formalism an arbitrary conformal superfield
is defined by the following OPEs ~\cite{NaKi}:
\ber
I^{a}(Z_{1})F^{\Lm}(Z_{2})=Z_{12}^{-1/2}E^{a}F^{\Lm}(Z_{2})+ reg., \nmb
I_{a}(Z_{1})F^{\Lm}(Z_{2})=Z_{12}^{-1/2}E_{a}F^{\Lm}(Z_{2})+ reg.,
\label{2.11}
\enr
here $E^{a}, E_{a}$ denote the generators of the ${\bf g^{\Bbb C}}$
in the representation with the highest weight $\Lm$,
\be
\Gm(Z_{1})F^{\Lm}(Z_{2})=Z_{12}^{-3/2}\Dl F^{\Lm}(Z_{2})+
Z_{12}^{-1}{1\ov 2} DF^{\Lm}(Z_{2})+
Z_{12}^{-1/2}\d F^{\Lm}(Z_{2})+ reg.,
\label{2.12}
\en
where the conformal dimension $\Dl$ is given by
\be
\Dl=C_{\Lm}/q,\
C_{\Lm}\equiv -(E^{a}E_{a}+E_{a}E^{a}),
\label{dim}
\en
and we have used the standard notations for even and odd world-sheet
super-intervals between a pair of points $Z_{i}=(z_{i},\Tta_{i}), i=1,2$:
\be
Z_{12}\equiv z_{1}-z_{2}-\Tta_{1}\Tta_{2},\
\Tta_{12}=Z^{1\ov 2}_{12}\equiv \Tta_{1}-\Tta_{2},
\label{2.13}
\en
so that
\be
Z_{12}^{n+{1\ov 2}}=Z_{12}^{n}\Tta_{12},\ n\in {\Bbb Z}.
\label{2.14}
\en

 We postulate the quantum version of the formula (\ref{tmap}):
\be
t: F^{\Lm}(Z) \to \brv{F}^{\Lm}(Z)= F^{\Lm}(Z)H(Z),
\label{qtmap}
\en
which is the quantum counterpart of (\ref{tmap})
(here and in what follows the leftmoving coordinate dependence of the fields
will be omitted for simplicity). It follows from the Sugawara formula (\ref{2.9})
and the OPEs (\ref{2.11}), (\ref{2.12}) that 
the conformal superfield $F^{\Lm}(Z)$ of the model satisfies 
the Knizhnik-Zamolodchikov equation ~\cite{NaKi}
\be
{q\ov 2}DF^{\Lm}(Z)+:F^{\Lm}I:(Z)=0,
\label{2.15}
\en
which is a quantization of the classical relation
$I=G^{-1}DG$. In view of (\ref{qtmap}) the dual field $\brv{F}^{\Lm}$ satisfies
the similar equation
\ber
{q\ov 2}D\brv{F}^{\Lm}(Z)=-:\brv{F}^{\Lm}\brv{I}:(Z)= \nmb
-:\brv{F}^{\Lm}H^{-1}IH:(Z)+{q\ov 2}\brv{F}^{\Lm}H^{-1}DH(Z).
\label{2.15d}
\enr

 Let us go back for a moment to the classical case and consider Eqs
(\ref{Jr4}), (\ref{Jr5}), and (\ref{H}). Using them we can write
\be
H^{-1}DH=2(\brv{I}^{+}-H^{-1}I^{-}H).
\en
As its quantum version we propose
\be
{q\ov 2}\brv{F}^{\Lm}H^{-1}DH(Z)= -2:\brv{F}^{\Lm}(\brv{I}^{+}-H^{-1}I^{-}H):(Z).
\label{2.16}
\en
The substitution (\ref{2.16}) converts (\ref{2.15d}) into
\be
:\brv{F}^{\Lm}(\brv{I}^{-}-\brv{I}^{+}):(Z)=
:\brv{F}^{\Lm}(H^{-1}(I^{+}-I^{-})H):(Z).
\label{2.17}
\en
Using the left-invariant complex structure $J$ on the group ${\bf G}$ one can rewrite
it in the form
\be
:\brv{F}^{\Lm}(JEnd(H)J\brv{I}):(Z)=:\brv{F}^{\Lm}I:(Z),
\label{2.18}
\en
where we have introduced the notation $End(H)x=HxH^{-1}, x\in {\bf g^{\Bbb C}}$
and we imply that $End(H)$ belongs to the group of super-Kac-Moody
algebra automorphisms. The equation (\ref{2.18}) means that $End(H)$ 
interchanges the isotropic subalgebras of the Manin triple because it
anticommutes with the complex structure $J$.
 
 By virtue of (\ref{2.18}) eq.(\ref{2.15d}) takes the form
\be
{q\ov 2}\brv{F}^{\Lm}H^{-1}DH(Z)=:\brv{F}^{\Lm}((End(H^{-1})JEnd(H)J-1)\brv{I}):(Z).
\label{2.19}
\en
Using super-Kac-Moody Ward identities ~\cite{NaKi} it is easy to see 
that (\ref{2.19}) decays into the system of equations
\ber
H^{-1}DH=0, \nmb
End(H^{-1})JEnd(H)J-1=0.
\label{2.20}
\enr
Its solution is given by the constant matrix anticommuting with $J$:
\ber
DH=0, \nmb
JEnd(H)+End(H)J=0.
\label{2.21}
\enr
In the orthonormal basis we have chosen, any matrix which anti-commutes with $J$ 
should have the form
\be
      \left(\begin{array}{cc}
            0      & 1\\
            1      & 0 \\ 
            \end{array}\right)
      \left(\begin{array}{cc}
            h &      0  \\
            0 & \bar{h} \\ 
            \end{array}\right),
\en
where $h$ is an arbitrary complex matrix 
(the bar denotes complex conjugation) .
Let us denote by $Aut({\bf g},J)$ the group of automorphisms of ${\bf g}$
which commute with $J$.
It is clear that 
\be
End(H)=\left(\begin{array}{cc}
            0      & 1 \\
            1      & 0 \\ 
            \end{array}\right)
\label{specH}
\en
is a solution of (\ref{2.21}). Hence each solution of 
(\ref{2.21}) should have the form:
\be
End(H)=\left(\begin{array}{cc}
            0      & 1 \\
            1      & 0 \\ 
            \end{array}\right)
       \left(\begin{array}{cc}
             m &       0 \\
             0 & \bar{m} \\ 
            \end{array}\right), \
       \left(\begin{array}{cc}
             m &       0 \\
             0 & \bar{m} \\ 
            \end{array}\right) \in Aut({\bf g},J).
\label{2.23}
\en
In view of (\ref{2.18}) $End(H)$ should be also an automorphism of the algebra 
$\hat{\bf g}$. It imposes on the matrix $m$ the relation
\be
m^{cb}\bar{m}_{ab}=\dlt^{c}_{a}.
\label{KMaut}
\en
The next condition we should demand is $t^{2}=1$ (that is, PL $T$-duality is an 
involution). It gives the second 
relation for $m$:
\be
m^{cb}\bar{m}_{ba}=\dlt^{c}_{a}.
\label{Inv}
\en
Therefore the set of PL $T$-duality transformations in the $N=2$ superconformal
WZNW model on the group manifold ${\bf G}$ is given by the set of matrices
(\ref{2.23}) satisfying (\ref{KMaut}), (\ref{Inv}).
Hence, under the quantum PL $T$-duality the currents (\ref{2.10})
transform as 
\be
t: I^{a}\to m^{ab}I_{b}, \
   I_{a}\to \bar{m}_{ab}I^{b},
\label{QImap}
\en
or in components,
\ber
t: \psi^{a}\to m^{ab}\psi_{b}, \ j^{a}\to m^{ab}j_{b}, \
   \psi_{a}\to \bar{m}_{ab}\psi^{b}, \ j_{a}\to \bar{m}_{ab}j^{b}.
\label{QImap1}
\enr
Taking into account (\ref{2.6}), (\ref{KMaut}), and (\ref{QImap1}) 
we find the PL $T$-duality transformation of the N=2 Virasoro 
superalgebra currents:
\ber
t: \Sgm^{\pm}\to \Sgm^{\mp}, \nmb
t: K\to -K, \ T\to T.
\label{QVirm}
\enr


 Notice that, as in the classical case, PL $T$-duality acts 
in the leftmoving sector as an identity transformation. Therefore
we may conclude that quantum PL $T$-duality in the $N=2$ superconformal
WZNW models is a mirror duality and has a geometric realization which is 
given by PL ${\bf G_{\pm}}$-holomorphic action on the target space of 
the model.

 Here a remark is in order. In many examples of the $N=2$ SWZNW models
on the compact groups ($SU(3)$, $SU(2)\times U(1)$, ...)
the transformations (\ref{QImap}) coincide with 
Weyl reflections. In these cases mirror symmetry was interpreted by the authors
of ~\cite{GiWi} as a gauge symmetry. They presented also a
contradictory example,
$SU(2)\times SU(2)$-SWZNW model, where 
the Weyl reflections failed to give mirror symmetry. It follows from 
our formula (\ref{QImap})
that in this example mirror symmetry is given by an external automorphism
of the Lie algebra $su(2)\times su(2)$. This example illustrates the general
picture: PL $T$-duality is given by an automorphism (internal or
external) which interchanges the isotropic subalgebras of the underlying Manin 
triple.


\vskip 10pt
\centerline{\bf3. Poisson-Lie $T$-duality and Mirror Symmetry in}
\centerline{\bf quantum Kazama-Suzuki models.}
 In this section we consider PL $T$-duality in Kazama-Suzuki models.
Kazama and Suzuki have studied  ~\cite{KaSu} the conditions under which
an N=1 superconformal coset model can have an extra supersymmetry,
giving rise to an $N=2$ superconformal model.
Then the $N=2$ superconformal coset
theories were classified more accurately in ~\cite{Schw}.
Their conclusion can be reformulated as follows.
Suppose the \MTR $({\bf g^{\Bbb C}},{\bf g_{+}},{\bf g_{-}})$
associated with the pair $({\bf g}, J)$ has a Manin subtriple 
$({\bf h},{\bf h_{+}},{\bf h_{-}})$,
that is, ${\bf h_{\pm}}\subset {\bf g_{\pm}}$ are subalgebras of ${\bf g_{\pm}}$
such that ${\bf h}\equiv {\bf h_{+}}\oplus {\bf h_{-}}$ is a subalgebra
of ${\bf g^{\Bbb C}}$ and $\tau: {\bf h_{+}}\to {\bf h_{-}}$.
Notice that the Manin subtriple specified above defines (with the help of the 
involution $\tau$) a pair $({\bf k},J)$ such that ${\bf k^{\Bbb C}}={\bf h}$
and ${\bf k}\subset {\bf g}$.

 Assume that the basis (\ref{Bas}) is chosen 
so that the subbases
\ber 
\{E^{i}, i=1,...,d_{h}\}, \nmb
\{E_{i}, i=1,...,d_{h}\}
\label{SBas}
\enr
are bases in the subalgebras ${\bf h_{+}}$  and ${\bf h_{-}}$, respectively.
Let us consider a vector subspace
\be
{\bf a}= {\bf g^{\Bbb C}}/ {\bf h}
\label{coset}
\en
generated (over ${\Bbb C}$) by the vectors
\be
\{E^{\al}, \al=d_{h}+1,...,d\}, \
\{E_{\al}, \al=d_{h}+1,...,d\}.
\label{csBas}
\en
The Manin triple construction of the Kazama-Suzuki models is given by the following

{\bf Proposition}. Suppose the isotropic subspaces
\be
{\bf a_{\pm}}\equiv{\bf a}\cap {\bf g_{\pm}}
\label{3.1}
\en
are Lie subalgebras. Then the currents
\ber
\Sgm^{+}_{cs}={2\over \sqrt{q}}(\psi^{\al}j_{\al}+
          {1\over 2}f_{\al \bt}^{\gm}:\psi^{\al}\psi^{\bt}\psi_{\gm}:),   \nmb
\Sgm^{-}_{cs}={2\over \sqrt{q}}(\psi_{\al}j^{\al}+
          {1\over 2}f^{\al\bt}_{\gm}:\psi_{\al}\psi_{\bt}\psi^{\gm}:),   \nmb
K_{cs}= ({2C^{\bt}_{\al}\over q} -\delta^{\bt}_{\al}):\psi^{\al}\psi_{\bt}:-
          {2\over q}(\hat{f}_{c}j^{c}-\hat{f}^{c}j_{c}),             \nmb
T_{cs}=-{1\over q}:(j^{a}j_{a}+j_{a}j^{a}):-
   {1\ov 2}:(\d \psi^{a}\psi_{a}-\psi^{a}\d \psi_{a}):+
   {1\ov q}:(u^{k}u_{k}+u_{k}u^{k}):,
\label{3.2}
\enr
where
\be
\hat{f}^{a}=f^{a\gm}_{\gm}, \ \hat{f}_{a}=f_{a\gm}^{\gm}, \
C^{\al}_{\bt}=\hat{f}^{a}f_{a\bt}^{\al}+ \hat{f}_{a}f^{a\al}_{\bt},
\label{3.3}
\en
\be
u^{k}=j^{k}-f^{k\al}_{\bt}:\psi^{\bt}\psi_{\al}:, \
u_{k}=j_{k}+f_{k\al}^{\bt}:\psi^{\al}\psi_{\bt}:,
\label{3.4}
\en
satisfy the OPEs of the $N=2$ super-Virasoro algebra with the central charge
\be
c_{cs}=c_{g}-c_{h}.
\label{3.5}
\en
This is just the $N=2$ extension ~\cite{HulS} of the GKO construction
formulated in terms of Manin triples and can be checked by direct 
calculations.

 The Kazama-Suzuki model based on the coset ${\bf G/K}$ can be obtained
from the SWZNW model on the group ${\bf G}$ by gauging an anomaly-free subgroup 
${\bf K}$ ~\cite{GawK}.
In view of the Manin triple construction (\ref{3.2}), (\ref{3.5})
this implies classically that the currents corresponding to the Manin subtriple
$({\bf h},{\bf h_{+}},{\bf h_{-}})$ should vanish: 
\be
I^{i}(Z)= I_{i}(Z)=0.
\en
In quantizing the theory canonically one should impose
in some way such constraints on physical states.
We impose
\ber
I^{i}(Z_{1})\Phi(Z_{2})= reg., \nmb
I_{i}(Z_{1})\Phi(Z_{2})= reg.,
\label{3.6}
\enr
that is, the physical states of the coset are the highest vectors
of the trivial $\hat{\bf h}$-representation.

 Under PL $T$-duality (\ref{QImap}) the set of constraints (\ref{3.6})
will transform, in general, into an other set of constraints giving another
coset model. Therefore we should define
PL $T$-duality transformations in the Kazama-Suzuki model as the subset
of (\ref{2.23})-(\ref{Inv}) which stabilizes the set (\ref{3.6}),
or equivalently, as the subset which stabilizes the Manin subtriple 
$({\bf h},{\bf h_{+}},{\bf h_{-}})$. Taking into account this condition
and using (\ref{QImap}) we obtain PL $T$-duality transformation rules
for the currents (\ref{3.2}) of the $N=2$ super-Virasoro algebra,
\ber
t: \Sgm^{\pm}_{cs}\to \Sgm^{\mp}_{cs}, \nmb
t: K_{cs}\to -K_{cs}, \ T_{cs}\to T_{cs},
\label{3.7}
\enr
which are similar to (\ref{QVirm}).
It is clear that PL $T$-duality in the leftmoving sector is given by the
identity transformation.

 Let us consider an example of the Kazama-Suzuki model based on the coset
$U(2)/(U(1)\times U(1))$ (the $N=2$ minimal model).
The complexification of $u(2)$ is the Lie algebra $gl(2,{\Bbb C})$.
In this case
the commutation relations (\ref{2.1}) in the orthonormal basis
(\ref{Bas}) are given by
\ber
\lbr E^{0},E^{1}\rbr=E^{1},     \nmb
\lbr E_{0},E_{1}\rbr=E_{1},     \nmb
\lbr E^{1},E_{1}\rbr=-E^{0}+E_{0}.
\label{3.8}
\enr
The isotropic subalgebras ${\bf g_{+}}$ and ${\bf g_{-}}$ of the complex Manin 
triple are generated by the vectors
$E^{0}, E^{1}$ and $E_{0}, E_{1}$ respectively.
The currents of the super-Kac-Moody algebra  $\hat{gl}(2,{\Bbb C})$ are
characterized by the following OPEs
\ber
I^{a}(Z_{1})I^{b}(Z_{2})= Z_{12}^{-1/2}f^{ab}_{c}I^{c}(Z)+reg., \nmb
I_{a}(Z_{1})I_{b}(Z_{2})= Z_{12}^{-1/2}f_{ab}^{c}I_{c}(Z)+reg., \nmb
I^{a}(Z_{1})I_{b}(Z_{2})= -Z_{12}^{-1}{q\ov 2}\dlt^{a}_{b} +
Z_{12}^{-1/2}(f_{bc}^{a}I^{c}-f^{ac}_{b}I_{c})+reg.,
\label{3.9}
\enr
where $a,b,c= 0,1$ and the structure constants are given by (\ref{3.8}).
The Manin subtriple defining our coset model is given by
\be
h=h_{+}\oplus h_{-}, \
h_{+}={\Bbb C}E^{0}, \
h_{-}={\Bbb C}E_{0}.
\label{Mstr}
\en
Thus, the Manin subtriple corresponds to the $N=2$ $U(1)^{2}$-SWZNW model which is
described by the pair of scalar complex free superfields $X^{0}(Z), X_{0}(Z)$ 
with obvious OPEs
\be
X^{0}(Z_{1})X_{0}(Z_{2})=-2\log{Z_{12}}.
\label{3.10}
\en
The currents of the super-Kac-Moody algebra $\hat{gl}(2,{\Bbb C})$ can be 
realized in terms of the fields $X^{0}(Z), X_{0}(Z)$ and
super-parafermions $S^{1}(Z), S_{1}(Z)$
~\cite{OhSuz}:
\ber
I^{0}={\sqrt{q}\ov 2}DX^{0}, \
I_{0}={\sqrt{q}\ov 2}DX_{0}, \nmb
I^{1}=\im S^{1}\exp{(-{1\ov \sqrt{q}}(X_{0}-X^{0}))}, \
I_{1}=\im S_{1}\exp{({1\ov \sqrt{q}}(X_{0}-X^{0}))}.
\label{3.11}
\enr
The super-parafermion OPEs
are deduced from the OPEs (\ref{3.9}), (\ref{3.10}) and the null-vector
relation in the trivial $\hat{su}(2)$-representation.

 The most general PL $T$-duality transformation in $U(2)$-SWZNW model
is given by
\ber
I^{0}\to I_{0}, \
I_{0}\to I^{0}, \nmb
I^{1}\to \exp{(\im \phi)}I_{1}, \
I_{1}\to \exp{(-\im \phi)}I^{1},
\label{3.12}
\enr
where $\phi$ is an arbitrary real number. We see that the constraints
transform into itself.
From these formulas we easily find the PL $T$-duality transformations
of the parafermions of the coset
\be
S^{1}\to \exp{(\im \phi)}S_{1}, \
S_{1}\to \exp{(-\im \phi)}S^{1}.
\label{3.13}
\en
Thus, the PL $T$-duality transformation acts in
the $U(1)^{2}$-subspace of the $U(2)$-SWZNW model as the usual 
$R\to 1/R$ $T$-duality
(at the self-dual point), while the PL $T$-duality 
transformation (\ref{3.13}) 
corresponds to the axial-vector duality of the coset $SU(2)/U(1)$ ~\cite{GiKir}
(to see this it is enough to recover the leftmoving
constraints).
 
 It is clear that there is a direct generalization of this example 
to the coset models ${\bf G/U(1)^{r}}$, where $r$ is the dimension of the
maximal torus of the group ${\bf G}$. The PL $T$-duality transformation
will act on the maximal torus as an Abelian $R\to 1/R$ $T$-duality
(at the self-dual point),
while in the N=2 Kazama-Suzuki model it will act as an axial-
vector duality ~\cite{Hen}. In the non-Abelian coset models 
the PL $T$-duality transformation rules of the fields are 
given by the non-Abelian generalization of the axial-vector
duality via ~\cite{Kir}. In principle they can be found using the non-Abelian 
generalization of the super-parafermions (\ref{3.11}).
Some aspects of this construction in the non supersymmetric case can be 
found in ~\cite{Lashk}.

Thus, in summary, we conclude that quantum PL T-duality in the 
Kazama-Suzuki models is a mirror duality also.

\vskip 10pt
\centerline{\bf4. Conclusion.}

 In this work we have considered the PL $T$-duality transformation in
quantum $N=2$ superconformal WZNW and Kazama-Suzuki models.
The PL $T$-duality transformation rules in the quantum $N=2$ SWZNW models
are found using the Manin triple construction of the $N=2$ SWZNW models,
the Knizhnik-Zamolodchikov equation, Ward identities, and
the conjecture that, as in the classical case, PL $T$-duality
is given by constant automorphisms of the rightmoving
super-Kac-Moody algebras of the models which interchange
the isotropic subalgebras of the underlying Manin triples. 
We have shown that in these models 
PL $T$-duality is a mirror duality. Thereby, we have thus given 
a geometric realization of the mirror symmetry in these models. 
Notice also that our results are in agreement with the conjecture 
proposed in ~\cite{GiWi} that mirror symmetry can be considered as a gauge 
symmetry (which is extended in some cases by the external automorphisms) of the
self-dual points of the moduli space of the $N=2$ SCFTs.

 We have given Manin triple construction of the Kazama-Suzuki models,
representing them as $(Manin \ triple/Manin \ subtriple)$-cosets.
By means of this representation we defined PL $T$-duality transformations
in the Kazama-Suzuki models as the subset of PL $T$-duality transformations
of the numerator triple which stabilize the denominator triple.
It was shown that, thus defined, PL $T$-duality is a mirror duality also.
An interesting open problem is to find the corresponding geometric
picture of PL $T$-duality and mirror symmetry in the classical 
Kazama-Suzuki models.

 Our results are useful in discussing Calabi-Yau superstring 
compactifications and allow us to conjecture that PL $T$-duality is
an adequate geometric structure underlying mirror symmetry.
The extension of our results to the Gepner construction of superstring vacua
~\cite{Gep} (see also ~\cite{Tao}) would be a test of the conjecture.

 Another interesting problem is to quantize the equations
(\ref{Jr4}), (\ref{Jr5}) and determine the quantum version of (\ref{42.1})
and (\ref{48}). Moreover, its solution is important in the context of
quantum PL $T$-duality and
mirror symmetry; it may be useful also in discussing $T$-duality for open 
strings and $D$-branes on curved backgrounds and will be helpful in
"quantization" of the existing treatments ~\cite{KlSWZ2}, ~\cite{KlSD}.

\vskip 10pt
\centerline{\bf ACKNOWLEDGMENTS}
\frenchspacing

This work was supported in part by grants
INTAS-95-IN-RU-690, CRDF RP1-277, RFBR 96-02-16507.

\vfill
\end{document}